\documentclass[amsmath,amssymb,aps,floatfix,twocolumn,pre,preprintnumbers,superscriptaddress]{revtex4-2}

\usepackage{graphicx}
\usepackage{upgreek}
\usepackage{amsmath}
\usepackage{amssymb}
\usepackage{chngpage}
\usepackage{tabularx}
\usepackage{array,booktabs}
\newcolumntype{C}{>{\centering\arraybackslash}X}
\usepackage{nicefrac}
\usepackage{color,soul}
\usepackage{xcolor}
\usepackage{enumitem}
\usepackage{relsize}

\begin{document}
	
	\title{Molecular dynamics simulations of $^1$H NMR relaxation in Gd$^{3+}$--aqua}

	\author{Philip M. Singer}
	\email{ps41@rice.edu}
	\affiliation{Department of Chemical and Biomolecular Engineering, Rice University, 6100 Main St., Houston, TX 77005, USA}
	\author{Arjun Valiya Parambathu}
	\affiliation{Department of Chemical and Biomolecular Engineering, Rice University, 6100 Main St., Houston, TX 77005, USA}
	\author{Thiago J. Pinheiro dos Santos}
	\affiliation{Department of Chemical and Biomolecular Engineering, Rice University, 6100 Main St., Houston, TX 77005, USA}
	\author{Yunke Liu}
	\affiliation{Department of Chemical and Biomolecular Engineering, Rice University, 6100 Main St., Houston, TX 77005, USA}
	\author{Lawrence B. Alemany}
	\affiliation{Shared Equipment Authority and Department of Chemistry, Rice University, 6100 Main St., Houston, TX 77005, USA}
	\author{George J. Hirasaki}
	\affiliation{Department of Chemical and Biomolecular Engineering, Rice University, 6100 Main St., Houston, TX 77005, USA}
	\author{Walter G. Chapman}
	\affiliation{Department of Chemical and Biomolecular Engineering, Rice University, 6100 Main St., Houston, TX 77005, USA}
	\author{Dilip Asthagiri}
	\email{dna6@rice.edu}
	\affiliation{Department of Chemical and Biomolecular Engineering, Rice University, 6100 Main St., Houston, TX 77005, USA}
	
	\begin{abstract}
		Atomistic molecular dynamics simulations are used to investigate $^1$H NMR $T_1$ relaxation of water from paramagnetic Gd$^{3+}$ ions in solution at 25$^{\circ}$C. Simulations of the $T_1$ relaxivity dispersion function $r_1$ computed from the Gd$^{3+}$--$^1$H dipole--dipole autocorrelation function agree within $\simeq 8$\% of measurements in the range $f_0 \simeq $ 5 $\leftrightarrow$ 500 MHz, without any adjustable parameters in the interpretation of the simulations, and without any relaxation models. The simulation results are discussed in the context of the Solomon-Bloembergen-Morgan inner-sphere relaxation model, and the Hwang-Freed outer-sphere relaxation model. Below $f_0 \lesssim $ 5 MHz, the simulation overestimates $r_1$ compared to measurements, which is used to estimate the zero-field electron-spin relaxation time. The simulations show potential for predicting $r_1$ at high frequencies in chelated Gd$^{3+}$ contrast-agents used for clinical MRI. 
	\end{abstract}
	
	\maketitle

	\section{Introduction}\label{sc:Introduction}
	
	The traditional theory of enhanced $^1$H NMR (nuclear magnetic resonance) relaxation of water due to paramagnetic transition-metal ions and lanthanide ions in aqueous solutions originates from Solomon \cite{solomon:pr1955}, Bloembergen and Morgan \cite{bloembergen:jcp1957,bloembergen:jcp1961}, a.k.a. the Solomon-Bloembergen-Morgan (SBM) model. The extended SBM model \cite{kowalewski:jcp1985} accounts for paramagnetic relaxation of inner-sphere water in the paramagnetic-ion complex \cite{solomon:pr1955,bloembergen:jcp1957,bloembergen:jcp1961}, outer-sphere water \cite{torrey:pr1953,hwang:JCP1975}, a.k.a. the Hwang-Freed (HF) model, the contact term \cite{bloembergen:jcp1957,morgan:jcp1959,bernheim:jcp1959,bloembergen:jcp1961}, the Curie term \cite{fries:jcp2003,helm:pnmrs2006}, and the electron-spin relaxation \cite{bloembergen:jcp1961,korringa:pr1962,westlund:book1995,fries:jcp2003,kowalewski:book,helm:pnmrs2006,belorizky:jcp2008}. The extended SBM model is most widely used in the interpretation of paramagnetic enhanced $^1$H relaxation of water due to Gd$^{3+}$-based contrast agents \cite{koenig:jcp1975,southwood:jcp1980,banci:1985,lauffer:cr1987,koenig:pnmrs1970,micskei:mrc1993,strandberg:jmr1996,powell:jacs1996,caravan:cr1999,rast:jcp01,bertini:book2001,borel:jacs2002,zhou:jmr2004,zhou:sap2005,yazyev:jcp2007,yazyev:ejic2008,lindgren:pccp2009,luchinat:jbio2014,aime:mp2019,fragai:cpc2019,li:jacs2019,washner:cr2019} used in clinical MRI (magnetic resonance imaging). The SBM model also forms the basis for the interpretation of paramagnetic relaxation in water-saturated sedimentary rocks \cite{kleinberg:jmr1994,foley:JMR1996,straley:sca2002,zhang:petro2003,korb:pre2009,mitchell:pnmrs2014,faux:pre2015,saidian:fuel2015}.
	
	The inner-sphere water constitutes the ligands of the Gd$^{3+}$ complex. The SBM inner-sphere model assumes a rigid Gd$^{3+}$--$^1$H dipole-dipole pair undergoing rotational diffusion, which according to the Debye model results in mono-exponential decay of the autocorrelation function. The Debye model was previously used in the Bloembergen-Purcell-Pound (BPP) model \cite{bloembergen:pr1948} for like spins (e.g. $^1$H--$^1$H pairs), and then adopted in the SBM  model for unlike spins (e.g. Gd$^{3+}$--$^1$H pairs). The SBM inner-sphere model also takes the electron-spin relaxation time into account, resulting in 6 free parameters for the $^1$H NMR relaxivity $r_1$. The inner-sphere relaxation is generally considered the largest contribution to relaxivity.
	
	The outer-sphere water are less tightly bound than the inner-sphere water. The HF outer-sphere model assumes that the Gd$^{3+}$ ion and H$_2$O are two force-free hard-spheres undergoing translational diffusion, which results in stretched-exponential decay of the autocorrelation function \cite{torrey:pr1953,hwang:JCP1975}. The HF outer-sphere model adds an additional 2 free parameters, bringing the total to 8 free parameters for the extended SBM model. Relaxation from the contact term is negligible for $^1$H NMR in Gd$^{3+}$--aqua \cite{powell:jacs1996,yazyev:jcp2007}, and is therefore neglected. Likewise, the Curie term is negligible in the present case \cite{fries:jcp2003,helm:pnmrs2006}.
	
	The application of the SBM and HF models to Gd$^{3+}$--aqua therefore requires fitting 8 free parameters over a large frequency range in measured $r_1$ dispersion (NMRD). In chelated Gd$^{3+}$ complexes, an order parameter plus a shorter correlation time \cite{lipari:jacs1982,lipari:jacs1982b} are added to the rotational motion of the complex \cite{fragai:cpc2019,aime:mp2019}, or to the electron-spin relaxation \cite{strandberg:jmr1996,zhou:sap2005,lindgren:pccp2009}, taking the total to 10 free parameters. This over-parameterized inversion problem often requires guidance in setting a range of values for the free parameters. It has also often speculated that the model for electron-spin relaxation is inadequate \cite{kowalewski:jcp1985}. 
	
	Atomistic molecular dynamics (MD) simulations can help elucidate some of these issues. MD simulation were previously used for like-spins, e.g. $^1$H--$^1$H dipole-dipole pairs, such as liquid-state alkanes, aromatics, and water 
	\cite{singer:jmr2017,asthagiri:seg2018,singer:jcp2018,asthagiri:jpcb2020}, as well as methane over a large range of densities \cite{singer:jcp2018b}. In all these cases, good agreement was found between simulated and measured $^1$H NMR relaxation and diffusion, without any adjustable parameters in the interpretation of the simulations. With the simulations thus validated against measurements, simulations can then be used to separate the intramolecular (i.e. rotational) from intermolecular (i.e. translational) contributions to relaxation, and to explore the corresponding $^1$H--$^1$H dipole-dipole autocorrelation functions in detail. For instance, MD simulations revealed for the first time ever that water and alkanes do not conform to the BPP model of a mono-exponential decay in the rotational autocorrelation function, except for highly symmetric molecules such as neopentane. More complex systems such as viscous polymers \cite{singer:jpcb2020} and heptane confined in a polymer matrix 
	\cite{parambathu:jpcb2020} have also been simulated, which again saw good agreement with measurements, and which lead to insights into the distribution in dynamic molecular modes.
	
	In this report, we extend these MD simulation techniques to Gd$^{3+}$--$^1$H dipole-dipole pairs, i.e. unlike spins, in a Gd$^{3+}$--aqua complex. The simulations show good agreement with measurements in the range $f_0 \simeq $ 5 $\leftrightarrow$ 500 MHz, without any adjustable parameters in the interpretation of the simulations, and without any relaxation models. These findings show potential for predicting $r_1$ at high frequencies in chelated Gd$^{3+}$ contrast agents used for clinical MRI. At the very least, the simulations could reduce the number of free parameters in the SBM and HF models, and help put constraints on its inherent assumptions for clinical MRI applications.

	\section{Methodology} \label{sc:Methodolgy}
	
	\subsection{Molecular simulation}\label{subsc:MD}
	
	To model the Gd$^{3+}$--aqua system, we use the AMOEBA polarizable force field to describe solvent water \cite{oldamoeba} and the ion \cite{Clavaguera:2006ee}. The Gd$^{3+}$ ion has an incomplete set of $4f$ orbitals, but this incomplete shell lies under the filled $5s$ and $5p$ orbitals. Thus ligand-field effects  \cite{Asthagiri:jacs04} are not expected to play a part in hydration and a spherical model of the ion ought to be adequate. However, polarization effects are expected to be important given the large charge on the cation. 
	
	Experimental NMR studies on Gd$^{3+}$--aqua use concentrations of about 0.3~mM. This amounts to having one GdCl$_3$ molecule in a solvent bath in a cubic cell about 176~{\AA} in size. Such large simulations are computationally rather demanding with the AMOEBA polarizable forcefield. 
	Further, at such dilutions, the ions essentially ``see'' only the water around them, and since understanding the behavior of water around the ion is of first interest, here we study a single Gd$^{3+}$ ion in a bath of 512 water molecules. (Note that within the Ewald formulation for electrostatic interactions, there is an implicit neutralizing background. This background does not impact the forces between the ion and the water molecules that are of interest here.) The partial molar volume of Gd$^{3+}$ in water at 298.15 K has been estimated by Marcus (1985) to be $-59.6$~cc/mol. We use this to fix the length of the cubic simulation cell to $24.805$~{\AA}. (From constant pressure simulations in a 2006 water system we find a Gd$^{3+}$ partial molar volume of about -63~cc/mol, in good agreement with the value suggested by Marcus. However, in this work, we will use the value suggested by Marcus.)

	All the simulations were performed using the OpenMM-7.5.1 package \cite{Eastman2017}. The van der Waals forces were switched to zero from 11~{\AA} to 12~{\AA}. The real space electrostatic interactions were cutoff at 9~{\AA} and the long-range electrostatic interactions were accounted using the particle mesh Ewald method with a relative error tolerance of 10$^{-5}$ for the forces. In the polarization calculations the (mutually) induced dipoles were converged to 10$^{-5}$~Debyes.
	
	The equations of motion were integrated using the “middle” leapfrog propagation algorithm with a time step of 1~fs coupled; this combination of method and time step provides excellent energy conservation in constant energy simulations. Exploratory work shows that the NMR relaxation in the Gd$^{3+}$-aqua system is sensitive to the system temperature. So we additionally use with a Nos{\'e}-Hoover chain \cite{nose:1984,hoover:1985} with three thermostats \cite{Zhang2019} to simulate the system at 298.15~K. The collision frequency of the thermostat was set to 100 fs to ensure canonical sampling. We carried out extensive tests with and without thermostats to ensure that the Nos\'e-Hoover thermostat does not affect dynamical properties. Our conclusions are consistent with an earlier study on good practices for calculating transport properties in simulations \cite{Maginn2018}.
	
	Initially, the system was equilibrated under $NVT$ conditions at 298.15 K for over 200~ps. We then propagated the system under the same $NVT$ conditions for 8~ns, saving frames every 0.1~ps for analysis. We used the last 6.5536~ns of simulation (equal to 65536($=2^{16}$) frames) for analysis. The mean temperature in the production phase was $298.15$~K with a standard error of the mean being 0.03~K. 
	
	\subsection{Structure and dynamics}
	
\begin{figure}[!ht]
		\begin{center}
			\includegraphics[width=1\columnwidth]{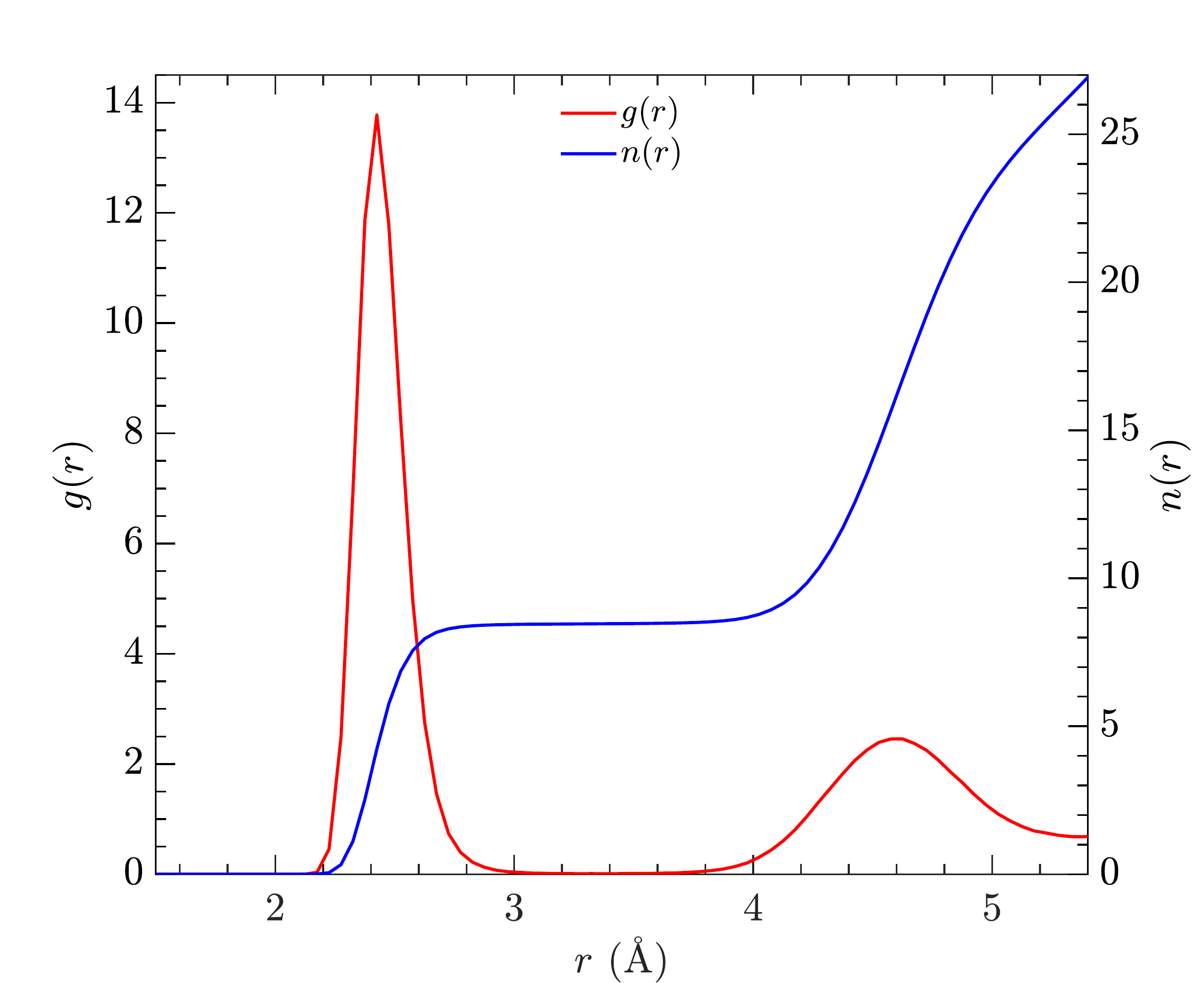}
		\end{center}
		\caption{Radial distribution function $g(r)$ of water oxygen around the Gd$^{3+}$ ion; the function $n(r)$ gives the coordination number.}
		\label{fg:Gr_Nr}
	\end{figure}

	Fig. \ref{fg:Gr_Nr} shows the ion-water radial distribution function. The location and magnitude of the peak is in good agreement with earlier studies founded on either \emph{ab initio} \cite{yazyev:jcp2007} or empirical force field-based \cite{lindgren:pccp2009} simulations. Consistent with those studies, we find that the first sphere (i.e. inner sphere) contains between $q = 8\leftrightarrow 9$ water molecules, with a mean of $q \simeq $ 8.5 consistent with published values \cite{koenig:jcp1975,banci:1985,powell:jacs1996,strandberg:jmr1996,rast:jcp01,lindgren:pccp2009,luchinat:jbio2014}. 
	
	\subsection{Residence time analysis}
	
	To estimate the residence time of water molecules in the inner sphere, we need to keep track of the water molecule as it moves into/out of the inner sphere. To this end, we follow the residence time of a defined water molecule $w$ using an indicator function $\chi_w$ that equals 1 if the water molecule is in the inner sphere and zero otherwise. The inner sphere is defined 
	as a sphere of radius $r \leq 3.3$~{\AA} around the Gd$^{3+}$ ion; this corresponds to the first hydration of the ion (Fig.~\ref{fg:Gr_Nr}). We perform this analysis for all the water molecules that visit the inner sphere at least once during the simulation. (We emphasize that the time here is discrete because configurations are saved only every 100~fs.) 
	
\begin{figure}[h!]
		\includegraphics[width=0.95\columnwidth]{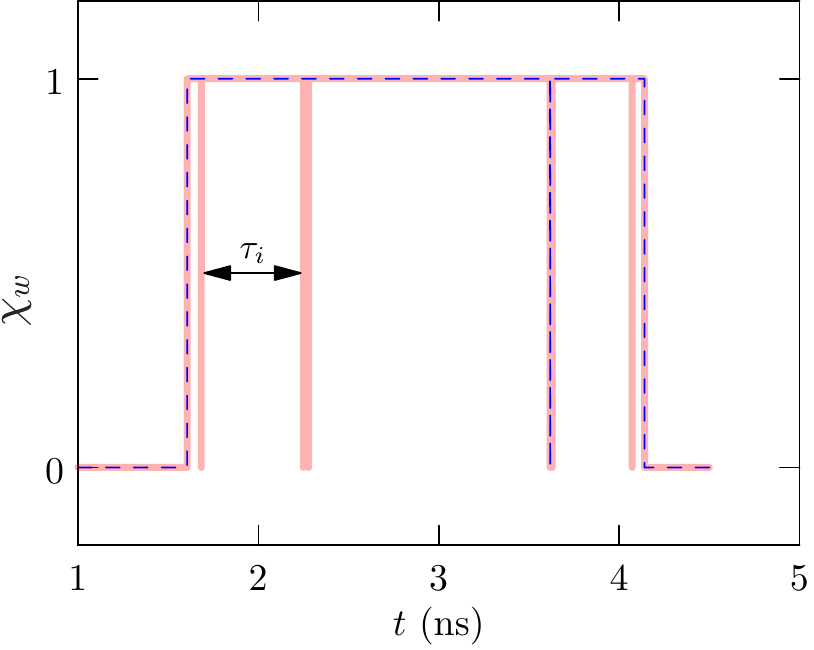}
		\caption{The trace of the indicator function $\chi_w$ for part of the simulation trajectory (red curve). Note the several transient escapes of the water molecule out of the inner sphere. The dashed (blue) curve is obtained by ``windowing" the raw data as noted in the text. $\tau$ is the length of time the water molecule spends continuously inside the inner sphere (one such domain, $\tau_i$, is shown).}
		\label{fg:restime}
	\end{figure}

	Fig.~\ref{fg:restime} shows the trace of the indicator function for a particular water molecule. 
	Note that during the approximately 4~ns window (from the $\approx 8$~ns trajectory), the water molecule makes several excursions out of the inner sphere before permanently leaving the inner sphere around 4.5~ns. The length of time that the water molecule spends continuously inside the inner sphere is denoted by $\tau_i$. As Fig.~\ref{fg:restime} shows, there can be several such islands of continuous occupation. 
	
	To test if the transient excursion is a bona fide escape from the inner sphere, we window average the data as follows: we consider a transient escape as a bona fide escape only if it persists for a defined number of consecutive time points. Fig.~\ref{fg:restime} shows the trace (blue curve) for such a ``windowing" for a window length of 200~fs, i.e.\ two consecutive frames in the trajectory of configurations. As can be expected, windowing extends the length of time that the molecule is defined to be inside the inner sphere. 
	
\begin{figure}[h!]
		\includegraphics[width=1\columnwidth]{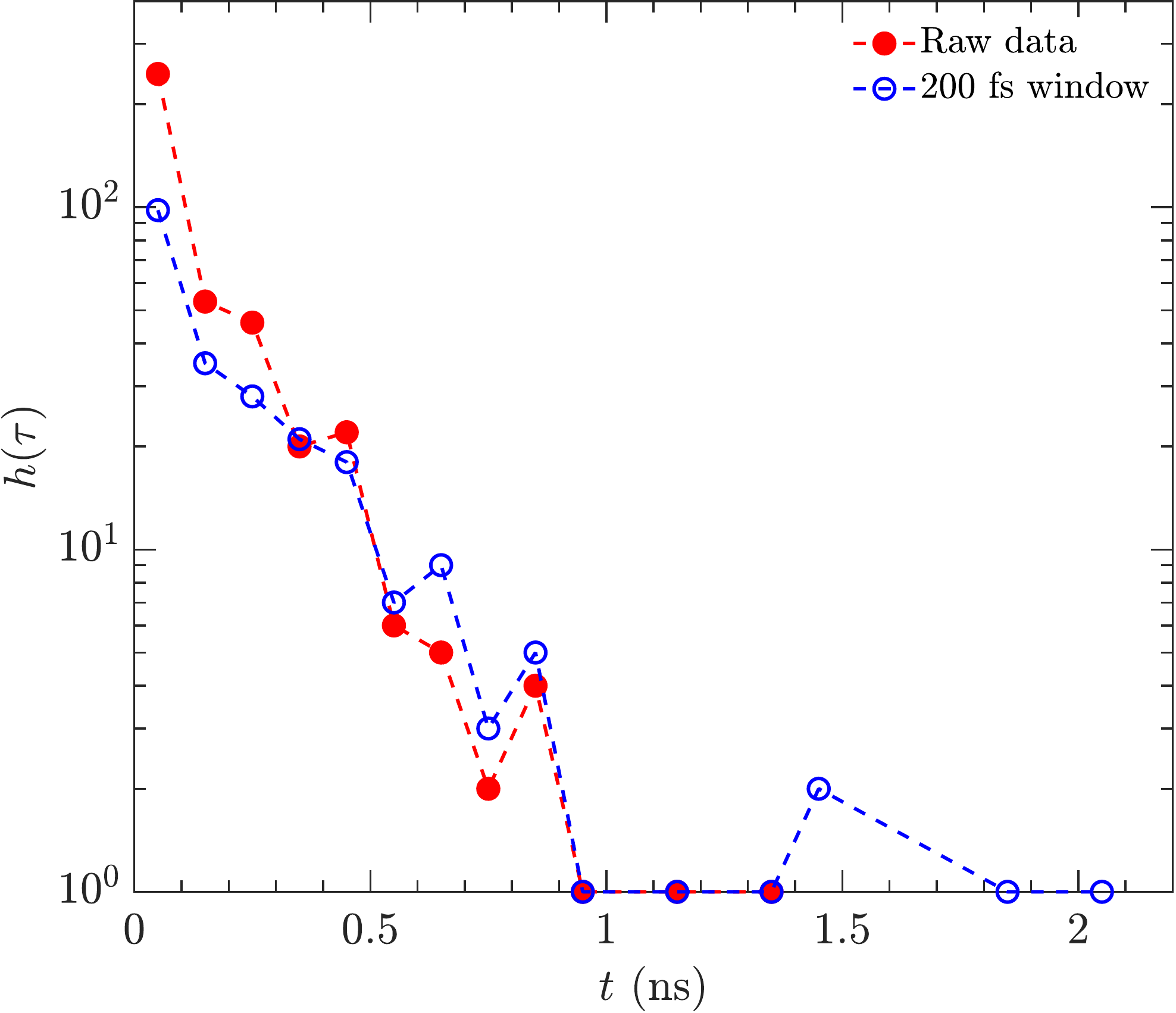}
		\caption{Distribution of continuous occupancy times $\tau$, $h(\tau)$, for raw data and 200 fs window.}
		\label{fg:residence}
	\end{figure}

	For all the water molecules that visit the inner sphere, we accumulate the set $\{\tau_i\}$ and then construct the histogram $h(\tau)$. Figure~\ref{fg:residence} shows the $h(\tau)$ for the raw data and the data with window length of 200~fs. 
	The $h(\tau)$ curve shows that there are many cases where water only transiently resides in the inner sphere. But we also find several water molecules that spend upwards of 0.5 ns continuously within the inner sphere. Specifically, for the data that has not been smoothed by ``windowing", we find three instances of water molecules continuously spending between 1 to 1.5~ns inside the inner sphere (and these also happen to be three distinct water molecules [data not shown]); with windowing using a 200~fs window, the upper limit is extended to just over  2~ns. Thus residence time, $\tau_m$, between $1 \leftrightarrow$ 2 ns are predicted for a complete rejuvenation of the inner sphere population. This time-scale is in accord with the range of published $^{17}$O NMR data that suggest residence times $\tau_m \simeq 1.0 \leftrightarrow 1.5$ ns \cite{southwood:jcp1980,powell:hel1993,micskei:mrc1993,helm:ccr99,borel:jacs2002}.
	
	\subsection{$^1$H NMR relaxivity} \label{subsc:Relaxation}
	
	The enhanced $^1$H NMR relaxation rate $1/T_{1}$ for water is given by the average over the $N = 1024$ $^1$H nuclei in the $L=24.805$ {\AA} box containing one paramagnetic Gd$^{3+}$ ion: 
	\begin{align}
		\frac{1}{T_{1}} &= \frac{1}{N}\sum_{i = 1}^{N}\frac{1}{T_{1i}}, \label{eq:R1} \\
		\frac{1}{T_{2}} &= \frac{1}{N}\sum_{i = 1}^{N}\frac{1}{T_{2i}}, \label{eq:R1_2} 
	\end{align}
	where $T_{1i}$ is the $T_1$ relaxation for the $i$'th $^1$H nucleus. The gadolinium molar concentration is given by $[Gd] =[H]/N$, where $[H] $ is the molar concentration of $^1$H in the simulation box. Equivalently, $[H] = 2[W]$ where $[W] = $ 55,705 mM is the molar concentration of H$_2$O at 25$^{\circ}$C. This leads to the following expression for the relaxivity in units of (mM$^{-1}$s$^{-1}$):
	\begin{align}
		r_1 &= \frac{1}{[Gd]} \frac{1}{T_1} = \frac{1}{[H]}\sum_{i = 1}^{N}\frac{1}{T_{1i}}, \label{eq:R2}\\
		r_2 &= \frac{1}{[Gd]} \frac{1}{T_2} = \frac{1}{[H]}\sum_{i = 1}^{N}\frac{1}{T_{2i}}. \label{eq:R2_2}
	\end{align}
	Note that $r_{1,2}$ are independent of $N$ (or box size $L$, equivalently). The ``fast-exchange" regime ($T_{1,2} \gg \tau_m$) is assumed \cite{korringa:pr1962}, and the chemical shift term in $r_2$ \cite{helm:pnmrs2006} is neglected for simplicity. Note also that the $^1$H--$^1$H dipole-dipole relaxation \cite{singer:jmr2017} is not considered in these simulations. 
	
\begin{figure}[!ht]
		\begin{center}	
			\includegraphics[width=1\columnwidth]{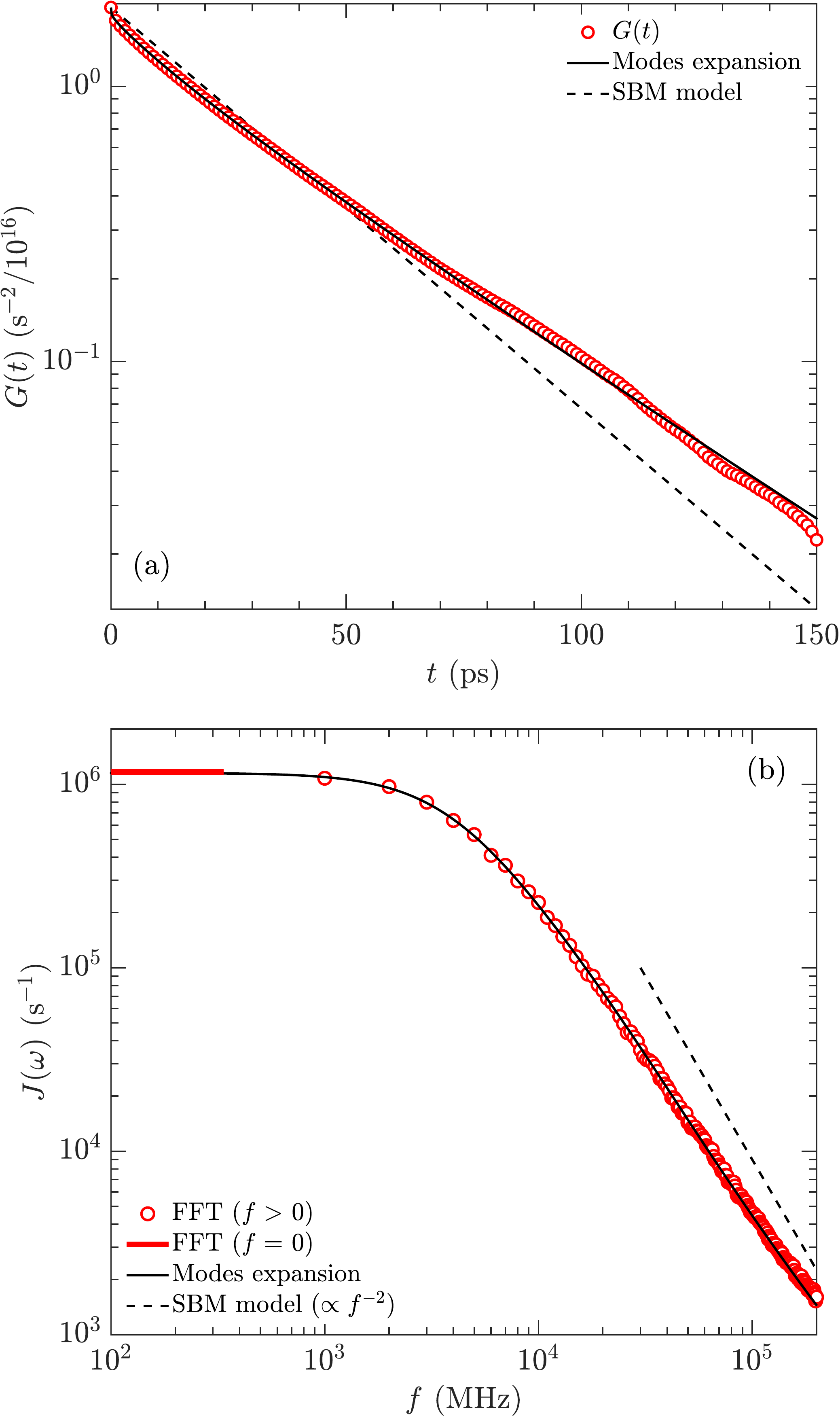} 	
		\end{center}
		\caption{(a) MD simulations of the autocorrelation function $G(t)$, where 1 in 10 data points are shown for clarity. Also shown is the modes expansion (Eq. \ref{eq:ILT1}) to $G(t)$, and SBM model (Eq. \ref{eq:HS1}) with $\tau_R = \left< \tau\right> $ = 30 ps. (b) Spectral density functions $J(\omega)$ from FFT (fast Fourier transform) (Eq. \ref{eq:R10}), including the $f = 0$ data point represented as a horizontal line placed at low frequency. Also shown is the modes expansion (Eq. \ref{eq:ILT6}), and the $f^{-2}$ power-law expected in the SBM model at high frequencies.}
		\label{fg:Gt_Jw}
	\end{figure}

	The computation of $T_{1i,2i}$ originates from the Gd$^{3+}$--$^1$H dipole-dipole autocorrelation function $G(t)$ \cite{bloembergen:pr1948,torrey:pr1953,abragam:book,mcconnell:book,cowan:book,kimmich:book} shown in Fig. \ref{fg:Gt_Jw}(a), where $t$ is the lag time. This autocorrelation is well suited for computation using MD simulations \cite{peter:jbnmr2001,case:acr2002}. Using the convention in the text by McConnell \cite{mcconnell:book}, $G_i(t)$ in units of (s$^{-2}$) is determined by:
	\begin{multline}
		G_i(t) = \frac{1}{4} \! \left(\frac{\mu_0}{4\pi}\right)^2 \! \hbar^2 \gamma_I^2\gamma_S^2 S(S+1) \times\\ 
		\left<\frac{(3\cos^{2}\!\theta_{i}\!(t+t')-1)}{r_{i}^3\!\left(t+t'\right)}  \frac{(3\cos^{2}\!\theta_{i}\!(t')-1)}{r_{i}^3\!(t')} \right>_{\!\! t'},
		\label{eq:R3}
	\end{multline}
	for the $i$'th $^1$H nucleus. $\theta_{i}$ is the angle between the Gd$^{3+}$--$^1$H vector ${\bf r}_{i} $ and the applied magnetic field ${\bf B}_0 $. $\mu_0$ is the vacuum permeability, $\hbar$ is the reduced Planck constant. $\gamma_I/2\pi = 42.58$ MHz/T is the nuclear gyro-magnetic ratio for $^1$H with spin $I = 1/2$, and $\gamma_S = 658\,\gamma_I$ is the electron gyro-magnetic ratio for Gd$^{3+}$ with spin $S = 7/2$. 
	
	Note that in Eq. \ref{eq:R3} we assume a spherically symmetric (i.e. isotropic) system, and therefore $G_i^m(t) = G_i(t)$ is independent of the order $m$, which amounts to saying that the direction of the applied magnetic field ${\bf B}_0 = B_0 \bf z$ is arbitrary. This assumption was verified in Ref. \cite{lindgren:pccp2009}. For simplicity, we therefore use the $m = 0$ harmonic $Y_2^0(\theta,\phi) = \sqrt{5/16\pi}\, (3\cos^{2}\theta-1) $ for the MD simulations \cite{singer:jmr2017}.
	
	The second-moment $\Delta\omega_i^2$ (i.e. strength) of the dipole-dipole interaction is defined as such \cite{cowan:book}:
	\begin{align}
		G_i(0) &= \frac{1}{3} \Delta\omega_i^2. \label{eq:R4}
	\end{align}
	Assuming the angular term in Eq. \ref{eq:R3} is uncorrelated with the distance term at $t = 0$, the relation  $\left<(3\cos^{2}\!\theta_{i}\!(\tau)-1)^2\right>_\tau = 4/5$ (which is independent of $i$) reduces the second moment to:
	\begin{align}
		\Delta\omega_i^2 = \frac{3}{5} \! \left(\frac{\mu_0}{4\pi}\right)^2 \! \hbar^2 \gamma_I^2\gamma_S^2 S(S+1)
		\left< \! \frac{1}{r_{i}^6(t')} \!\right>_{\!\! t'}.
		\label{eq:R5}
	\end{align}
	
	The next step is to take the (two-sided even) fast Fourier transform (FFT) of the $G_i(t)$ to obtain the spectral density function:
	\begin{align}
		J_i(\omega) &= 2\int_{0}^{\infty} \! G_i(t) \cos(\omega t)\, dt. \label{eq:R10}
	\end{align}
	The relaxation rates are then determined for unlike spins \cite{mcconnell:book}:
	\begin{align}
		\frac{1}{T_{1i}} &= J_i(\omega_0)  + \frac{7}{3} J_i(\omega_e), \label{eq:R11} \\
		\frac{1}{T_{2i}} &= \frac{2}{3}J_i(0)  + \frac{1}{2}J_i(\omega_0)  + \frac{13}{6} J_i(\omega_e), \label{eq:R12}
	\end{align}
	assuming $\omega_e \gg \omega_0$, where $\omega_0 = \gamma_I B_0 = 2\pi f_0$ is the $^1$H NMR resonance frequency, and $\omega_e = 658\,\omega_0$ is the electron  resonance frequency.
	
	The expressions for $T_{1i,2i}$ are then summed in Eqs. \ref{eq:R2} and \ref{eq:R2_2} to compute $r_1$ and $r_2$, respectively. We also define the following quantities summed over the $N = 1024$ $^1$H nuclei:
	\begin{align}
		G(t) &=  
		\sum\limits_{i = 1 }^{N} \! G_i(t),
		\label{eq:R7}\\
		\Delta\omega^2 &=\sum\limits_{i = 1 }^{N} \Delta\omega_i^2 \label{eq:R8}\\
		J(\omega) &=\sum\limits_{i = 1 }^{N} J_i(\omega). \label{eq:R9}
	\end{align}
	Note that the summed quantities $G(t)$, $\Delta\omega^2$, and $J(\omega)$ are independent of $N$ (or box size $L$, equivalently). The $G(t)$ simulation data is plotted in Fig \ref{fg:Gt_Jw}(a), while the $J(\omega)$ simulation data is plotted in Fig \ref{fg:Gt_Jw}(b). 
	
	We also define the average correlation time $\left<\tau\right>$ as the normalized integral \cite{cowan:book}:
	\begin{align}
		\left<\tau\right> &= \frac{1}{G(0)}\int_{0}^{\infty}\! G(t)\,dt. \label{eq:tau} 
	\end{align}
	The low frequency (i.e. extreme narrowing) limit $r_1(0) = r_2(0)=r_{1,2}(0)$ can then be expressed as:
	\begin{align}
		r_{1,2}(0)  =\frac{1}{[H]}\frac{20}{9} \Delta\omega^2\!\left<\tau\right>. \label{eq:R15}
	\end{align}
	Note how $r_{1,2}(0)$ (for unlike Gd$^{3+}$--$^1$H spin pairs) is a factor 2/3 less than the equivalent expression for like $^1$H--$^1$H spin pairs \cite{singer:jmr2017}. 
	
	We assume the ``fast-exchange" regime in the above formulation, which is discussed in more detail in Section \ref{subsc:SBM}. The fast-exchange regime can be inferred directly from measurements since $r_1$ increases with decreasing temperature \cite{koenig:jcp1975}. Investigations are underway to extend the simulations to the slow-exchange regime.
	
	The above analysis also assumes the electron spin is a point-dipole centered at the Gd$^{3+}$ ion \cite{kowalewski:jcp1985}. Given that the simulation agrees with measurements in the range $f_0 \simeq $ 5 $\leftrightarrow$ 500 MHz, the point-dipole approximation is considered valid in the present case for $^1$H \cite{helm:pnmrs2006}.

	\subsection{Expansion of $G(t)$ in terms of molecular modes}\label{subsc:ILT}
	The FFT result for $J(\omega)$ in Fig. \ref{fg:Gt_Jw}(b) is sparse. Besides the $f$ = 0 data point, the lowest frequency FFT data point is given by the resolution $\Delta f = 1/2t_{max} = $ 10$^3$ MHz, where $t_{max} = 500$ ps is the longest lag time in $G(t)$. 
	
\begin{figure}[!ht]
		\begin{center}
			\includegraphics[width=1\columnwidth]{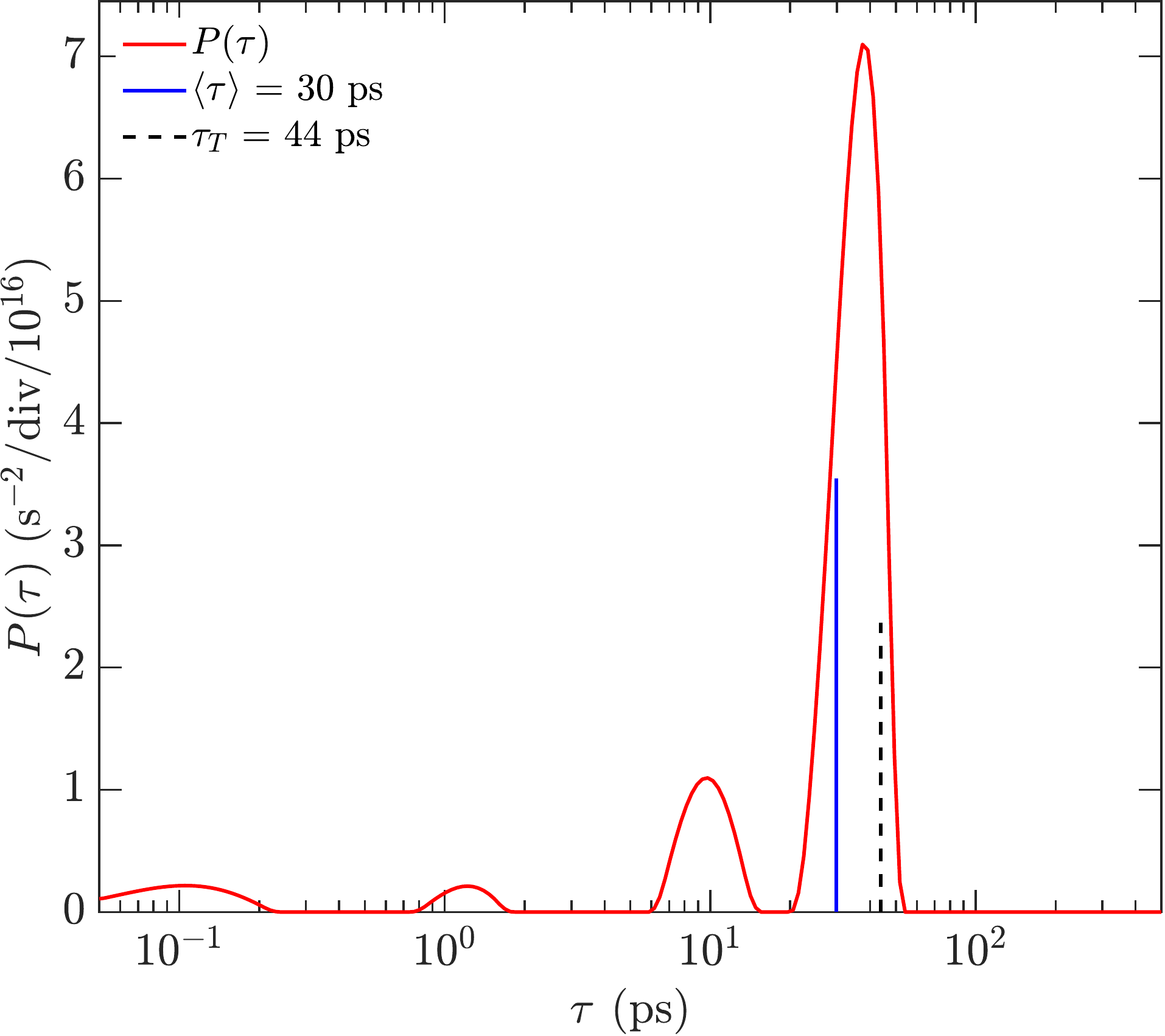} 			
		\end{center}
		\caption{Probability density function $P(\tau)$ obtained from the expansion of molecular modes (Eq. \ref{eq:ILT1}) of $G(t)$. The average correlation time $\left<\tau\right>$ (Table \ref{tb:MD}) and predicted translational correlation time $\tau_T$ (Table \ref{tb:Deff}) are shown.}
		\label{fg:Ptau}
	\end{figure}

	As an alternative to zero-padding the FFT,  we see to model $G(t)$ in terms of molecular modes. To this end, 
	we expand $G(t)$ as 
	\begin{align}
		G(t) &= \int_{0}^{\infty}\! P(\tau) \exp\left(-\frac{t}{\tau}\right) d\tau, 
		\label{eq:ILT1} 
	\end{align}
	where $P(\tau)$ is the underlying distribution in molecular correlation times, $\tau$.  We solve this Fredholm integral equation of the first kind to recover the $P(\tau)$ distribution.  Since $G(t)$ is available only at discrete time intervals, the inversion is an 
	ill-posed problem. We address this by using Tikhonov regularization \cite{singer:jcp2018,singer:prb2020}, with the vector $\textbf{P}$ being one for which 
	\begin{eqnarray}
		{\bf P} = \underset{{\bf P}\geq0}{\mathrm{arg\, min}}\,\,  || {\bf G} - K \,{\bf P}||^2 + \alpha ||{\bf P}||^2   \label{eq:ILT4}
	\end{eqnarray}
	is a minimum. Here $\textbf{G}$ is the column vector representation of the autocorrelation function $G(t)$, $\textbf{P}$ is the column vector representation of the distribution function $P(\tau)$, $\alpha$ is the regularization parameter, and $K$ is the kernel matrix:
	\begin{align}
		K = K_{ij} = \exp\left(-\frac{t_i}{\tau_{j}}\right). \label{eq:ILT5}
	\end{align}
	The results for $P(\tau)$ are shown in Fig. \ref{fg:Ptau}, from which the following are determined:
	\begin{align}
		\left<\tau\right> &= \frac{1}{G(0)}\int_{0}^{\infty}\! P(\tau)\, \tau \,d\tau, \label{eq:ILT2} \\
		G(0) &= \int_{0}^{\infty}\! P(\tau) d\tau= \frac{1}{3} \Delta\omega^2. \label{eq:ILT3} 
	\end{align}
	The spectral density $J(\omega)$ is then determined from the Fourier transform (Eq. \ref{eq:R10}) of $G(t)$ (Eq. \ref{eq:ILT1}): 
	\begin{align}
		J(\omega) &= \int_{0}^{\infty}\! \frac{2\tau}{1+(\omega\tau)^2}   P(\tau) d\tau \, ,\label{eq:ILT6}
	\end{align}
	from which $T_{1,2}$ at any desired $f_0$ can be determined (Eqs. \ref{eq:R11} and \ref{eq:R12}). More in-depth discussions of the above procedure, loosely termed as ``inverse Laplace transform", can be found in Refs. \cite{parambathu:jpcb2020,singer:jpcb2020,singer:jcp2018b,asthagiri:jpcb2020,wang:prb2021,imai:jpsj2021} and the supplementary material in Refs.~\cite{singer:jcp2018,singer:prb2020}. We hasten to add that inverting Eq.~\ref{eq:ILT1} to recover $P$ is formally not a Laplace inversion \cite{fordham:diff2017}, but this terminology is common in the literature. 
	Possible alternatives to the above formulation are also discussed in Ref. \cite{asthagiri:jpcb2020}.
	
	The $P(\tau)$ is binned from $\tau_{min} = 0.05$ ps to  $\tau_{max} = 500$ ps using 200 logarithmically spaced bins. In the present case of low-viscosity fluids ($\eta \simeq $ 1 cP), the choice of $\tau_{max} $ does not impact $J(\omega)$, and is therefore {\it not} a free parameter in the analysis in terms of molecular modes. The constant ``div" in Fig. \ref{fg:Ptau} is a ``division" on a log-scale. More specifically, div $= \log_{10}(\tau_{i+1}) - \log_{10}(\tau_{i})$ is independent of the bin index $i$, and ensures unit area when $P(\tau)$ is of unit height and a decade wide \cite{singer:prb2020}. 
	
	As shown in Fig. \ref{fg:Gt_Jw}(a), the residual between the $G(t)$ data and the fit using molecular modes is not dominated by Gaussian noise. As such, the regularization parameter is fixed to $\alpha =$ 10$^{-1}$ in accordance with previous studies \cite{singer:jcp2018,singer:jcp2018b,singer:jpcb2020,parambathu:jpcb2020,asthagiri:jpcb2020}. As shown in Fig. \ref{fg:Gt_Jw}(b), we find that $\alpha =$ 10$^{-1}$ gives excellent agreement with the parameter-free $J(\omega)$ from FFT, which validates the analysis in terms of molecular modes. The results in Fig. \ref{fg:Gt_Jw}(b) further emphasize the following advantages: 
	(1) the expansion (Eq.~\ref{eq:ILT1}) filters out the noise while still honoring the FFT data (including the $f = 0$ data point), (2) Eq.~\ref{eq:ILT6} provides $J(\omega)$ for any desired $f$ value, and (3) the expansion in terms of molecular modes
	leads to physical insight into the distribution $P(\tau)$ of molecular correlation times $\tau$.

	\subsection{Diffusion}\label{subsc:Diffusion}
	
	An independent computation of translational diffusion $D_T$ was performed from MD simulations. We calculate the mean square displacement $\left<\!\Delta r^2\right>$ of the water oxygen and Gd$^{3+}$ ion as a function of the diffusion evolution time $t$ ($<$ 10 ps), and additionally average over a sample of 50 molecules to ensure adequate statistical convergence.  
	\begin{figure}[h!]
		\begin{center}
			\includegraphics[width=1\columnwidth]{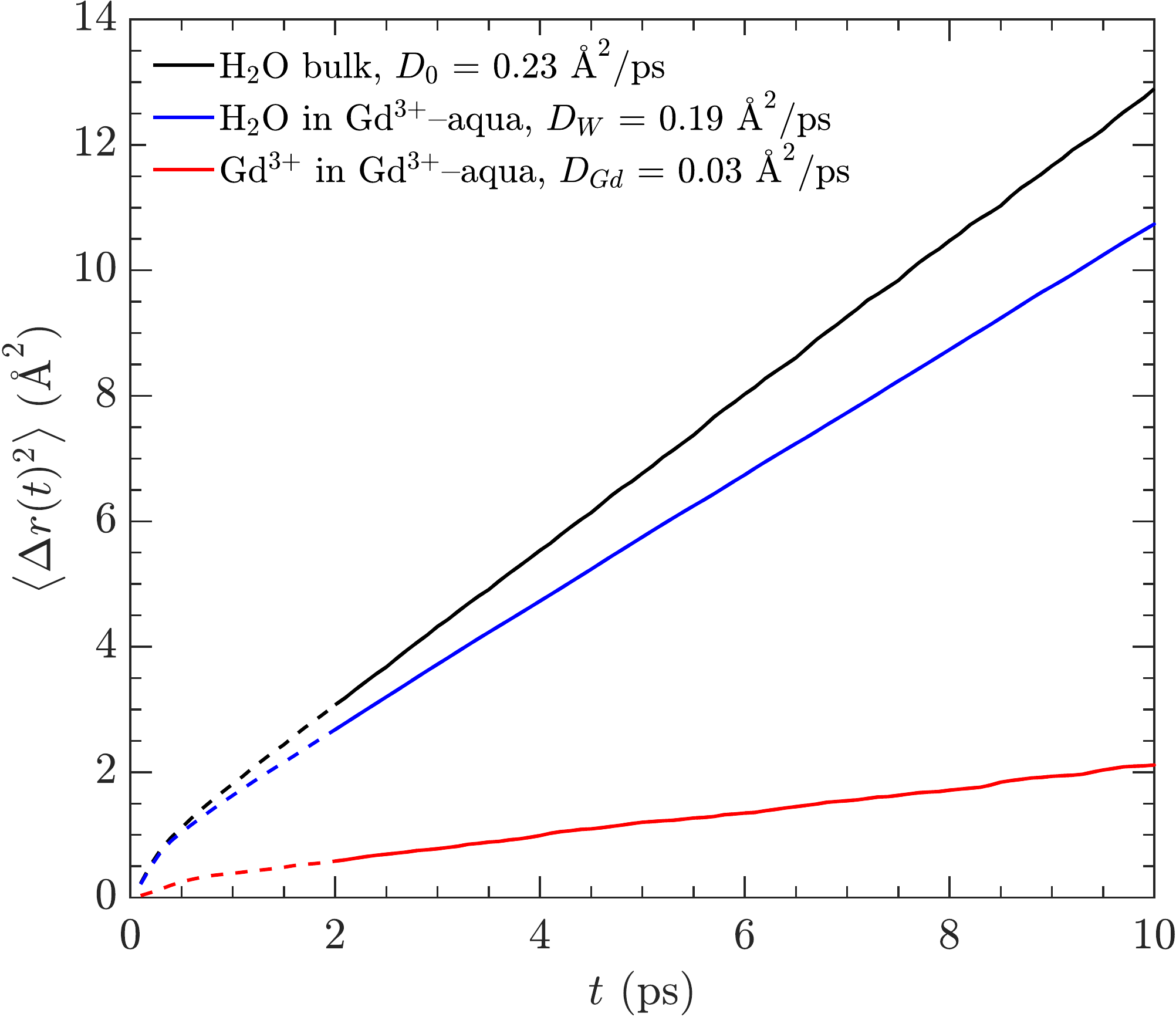} 
		\end{center}
		\caption{MD simulations of mean-square displacement $\left<\!\Delta r^2 \right>$ versus time $t$ for bulk water, water in Gd$^{3+}$--aqua, and Gd$^{3+}$ in Gd$^{3+}$--aqua. Solid lines show fitting region used to obtain translational diffusion coefficient $D_{sim}$ from Eq. \ref{eq:D1} for $t > 2 $ ps, and dashed lines show early time regime not used in the fit. Legend indicates $D_T$ values which include the correction term (Eq. \ref{eq:D2}).} 
		\label{fg:Diff}
	\end{figure}
	
	As shown in Fig. ~\ref{fg:Diff}, at long-times ($t$) the slope of the linear diffusive regime gives the translational self-diffusion coefficient $D_{sim}$ according to Einstein's relation:
	\begin{equation}
		D_{sim} = \frac{1}{6} \frac{\delta \! \left<\!\Delta r^2\right>}{\delta t}  \, 
		\label{eq:D1}
	\end{equation}
	where $D_{sim}$ is the diffusion coefficient obtained in the simulation using periodic boundary conditions in a cubic cell of length $L$. In the linear regression procedure, the early ballistic regime and part of the linear regime is excluded to obtain a robust estimate of $D_{sim}$.
	
	Following Yeh and Hummer \cite{yeh:jpcb2004} (see also D\"unweg and Kremer \cite{kremer:jcp93}), we obtain the diffusion coefficient for an infinite system $D_{T}$ from $D_{sim}$ using 
	\begin{equation}
		D_{T} = D_{sim} +  \frac{k_B T}{6 \pi \eta} \frac{\xi}{L} \label{eq:D2}
	\end{equation}
	where $\eta$ is the shear viscosity and $\xi = 2.837297$ \cite{yeh:jpcb2004} is the same quantity that arises in the calculation of the Madelung constant in electrostatics. (In the electrostatic analog of the hydrodynamic problem, $\xi/L$ is the potential at the charge site in a Wigner lattice.) For the system sizes considered in this study, $L \simeq 25$~{\AA}, the correction factor constituted $\simeq13\%$ of $D_0$, and $\simeq16\%$ of $D_W$. The correction factor was not applied to $D_{Gd}$.
	
	\subsection{Measurements} \label{subsc:measurements}
	We prepared a Gd$^{3+}$--aqua solution in de-ionized water at $[Gd]$ = 0.3 mM and measured $T_{1,meas}$ at a controlled temperature of 25$^{\circ}$C, using static fields at $f_0 = 2.3$ MHz with an Oxford Instruments GeoSpec2, at $f_0 = 20$ MHz with a Bruker Minispec, and at $f_0 = 500$ MHz with a Bruker 500 MHz Spectrometer. The measured relaxivity $r_1$ was determined as such:
	\begin{align}
		r_1 = \frac{1}{[Gd]}\left(\!\frac{1}{T_{1,meas}} - \frac{1}{T_{1,bulk}}\!\right). \label{eq:M1}
	\end{align}
	where $T_{1,bulk} = $ 3.13 s was found for bulk water (not de-oxygenated \cite{shikhov:amr2016}) at 2.3 MHz and 500 MHz. Field cycling $r_1$ data at $[Gd]$ = 1 mM and 25$^{\circ}$C were taken from Luchinat {\it et al.} \cite{luchinat:jbio2014} (supplementary material) using a Stelar SpinMaster 1T. The field cycling results agreed with our measurements at $f_0 = $ 2.3 MHz and 20 MHz (within $\simeq$ 5\%), while our $f_0 = 500$ MHz significantly extends the frequency range of the measurements.

	\section{Results and discussions} \label{sc:Results}
	
	In this section we compare the simulated relaxivity $r_1$ with measurements. The simulated relaxivity is then discussed in the context of the Solomon-Bloembergen-Morgan (SBM) model, and Hwang-Freed (HF) model. The zero-field electron-spin relaxation time is then determined from $r_1$ at low frequencies. 
	
	\subsection{Comparison with measurements}
	
\begin{figure}[!ht]
		\begin{center}
			\includegraphics[width=1\columnwidth]{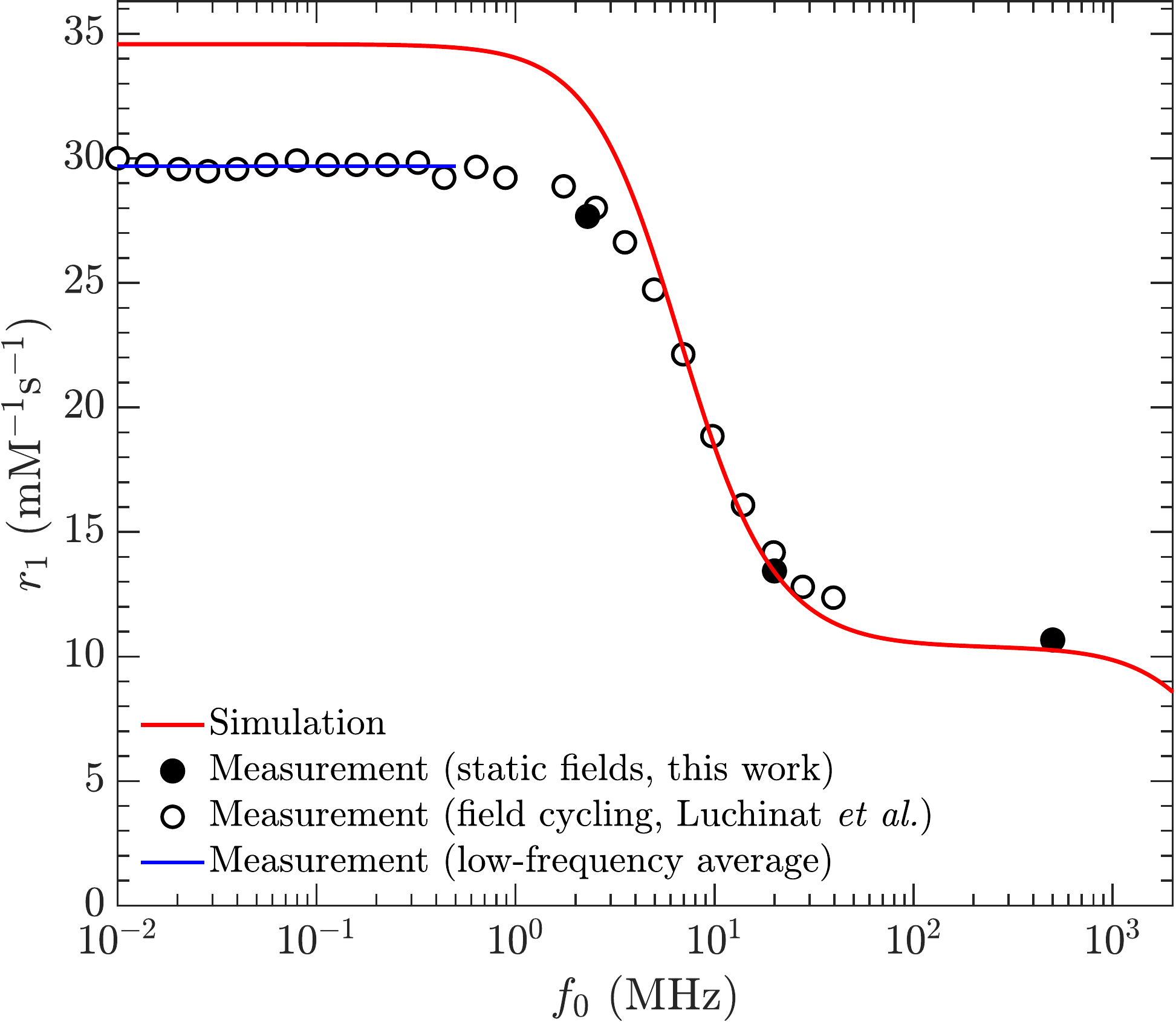} 
		\end{center}
		\caption{Simulated $^1$H NMR relaxivity $r_1$ of Gd$^{3+}$--aqua solution at 25$^{\circ}$C, compared with static-field measurements (this work) and field-cycling measurements (Luchinat {\it et al.} \cite{luchinat:jbio2014}, supplementary material). Also shown is the average of the low-frequency ($f_0<0.5$ MHz) measurements.}
		\label{fg:R1_relaxivity}
	\end{figure}

\begin{figure}[!ht]
		\begin{center}
			\includegraphics[width=0.9\columnwidth]{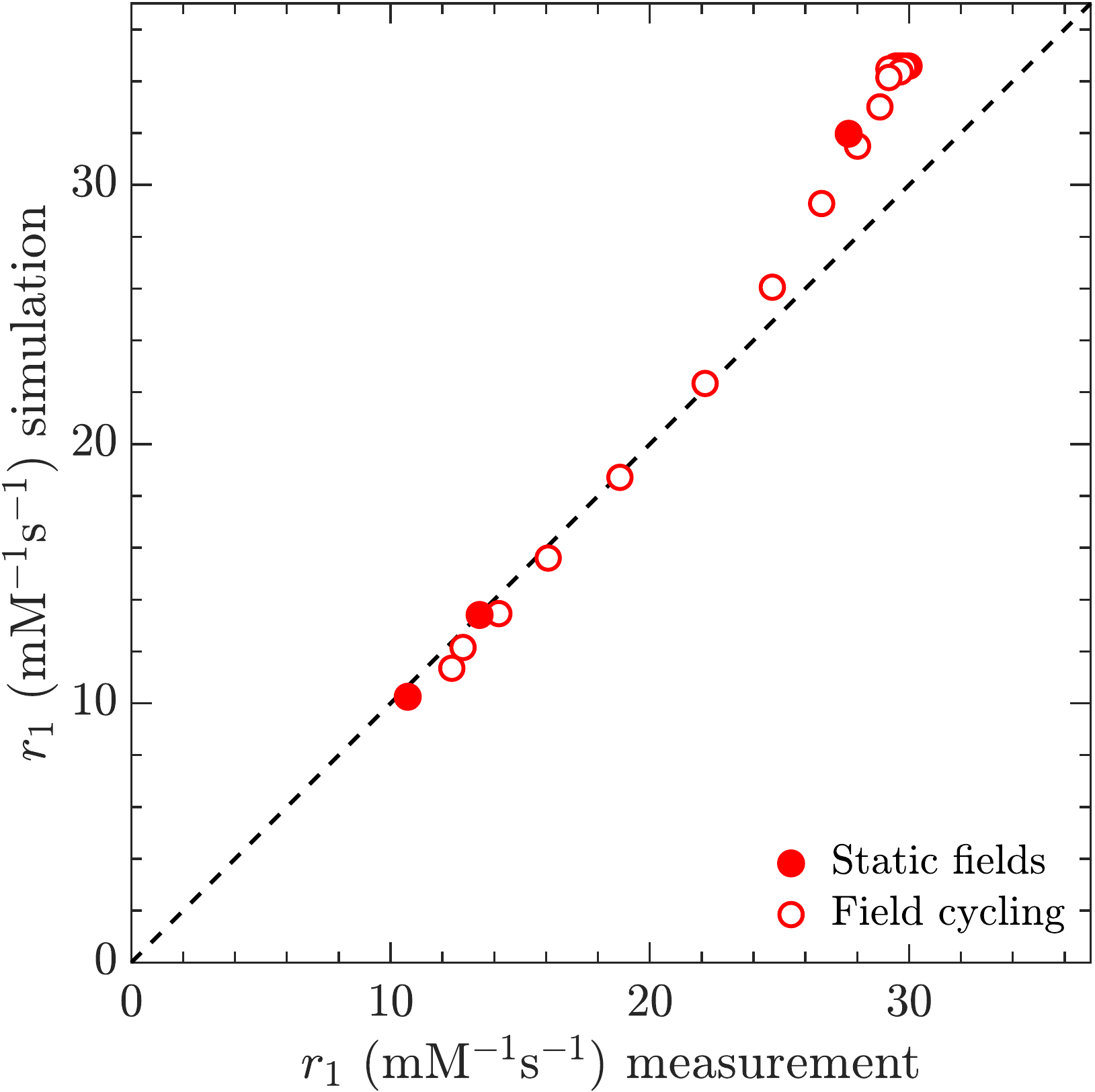} 
		\end{center}
		\caption{Cross-plot of measured $r_1$ (including static-field and field-cycling measurements) versus simulations taken from Fig. \ref{fg:R1_relaxivity}, all at 25$^{\circ}$C. Dashed line is the 1-1 unity line.}
		\label{fg:R1_cross}
	\end{figure}

	The results of simulated relaxivity $r_1$ (Eq. \ref{eq:R2}) are shown in Fig. \ref{fg:R1_relaxivity}, alongside corresponding measurements of Gd$^{3+}$--aqua solution at 25$^{\circ}$C using field cycling by Luchinat {\it et al.} \cite{luchinat:jbio2014} (supplementary material), and static fields from this work. A cross-plot of simulated versus measured $r_1$ results are also shown in Fig. \ref{fg:R1_cross}. The simulation is within $\simeq$ 8\% of measurements in the range $f_0 \simeq $ 5 $\leftrightarrow$ 500 MHz. Given that there are no adjustable parameters in the interpretation of the simulations, this agreement validates our simulations at high frequencies. We note that a similar degree of agreement (within $\simeq$ 7\%) was found in previous studies of liquid alkanes and water \cite{singer:jmr2017}. 
	
	For convenience, Table \ref{tb:MD} lists the average correlation time $\left<\tau\right>$ (Eq. \ref{eq:ILT2}), the square-root of the second moment (i.e. strength) of $\Delta \omega$ (Eq. \ref{eq:ILT3}), and the residence time $\tau_m$ (Fig. \ref{fg:residence}). Note that these three quantities are model free.
	
	\begin{table}[!ht]
		\centering
		\begin{tabular}{ccc|cccc|c}				
			\hline
			$^{{\strut}{}}$	$\left<\tau\right>$&  $\Delta \omega/2\pi$&  $\tau_m$ &$q$ &  $r_{in}$  &  $\Delta \omega_{in}/2\pi$& $T_{1in}(0)$ & $T_{e0}$\\
			$^{{\strut}{}}$	(ps) & (MHz)& (ns)& & ({\AA})& (MHz)& ($\mu$s) &(ps) \\
			\hline
			$^{{\strut}{}}$	30 & 38.4 & 1 $\leftrightarrow$ 2 & 8.5 & 2.97 & 9.3 & 4.4 & 180
		\end{tabular}
		\caption{Analysis of the simulation results including; (left) mean correlation time $\left<\tau\right>$ (Eq. \ref{eq:ILT2}), square-root of second moment $\Delta\omega$ (Eq. \ref{eq:ILT3}), and residence time $\tau_m$ (Fig. \ref{fg:residence}); (middle) approximate inner-sphere quantities including coordination number $q$ (Fig. \ref{fg:Gr_Nr}), Gd$^{3+}$--$^1$H distance $r_{in}$ (Eq. \ref{eq:SBM_R}), square-root of second moment $\Delta\omega_{in}$ (Eq. \ref{eq:SBM_SM}), and relaxation time $T_{1in}(0)$ at $f_0 = 0$ (Eq. \ref{eq:SBM_T1}); (right) zero-field electron-spin relaxation time $T_{e0} $ (Eq. \ref{eq:EL2}).}\label{tb:MD}
	\end{table}

	\subsection{SBM inner-sphere model}\label{subsc:SBM}
	
	The SBM inner-sphere model assumes a rigid Gd$^{3+}$--$^1$H pair undergoing rotational diffusion, leading to the following mono-exponential decay in the autocorrelation function:
	\begin{align}
		G_{SBM}(t) = G_{SBM}(0)\exp\left(-\frac{t}{\tau_R} \right). \label{eq:HS1}
	\end{align}
	This functional form is identical to the BPP model \cite{bloembergen:pr1948} which is based on the Debye model, where the rotational-diffusion correlation time $\tau_R$ is defined as the average time it takes the rigid pair to rotate by 1 radian. $G_{SBM}(t)$ is plotted in Fig. \ref{fg:Gt_Jw}(a) assuming $\tau_R = \left<\tau\right>$ = 30 ps. 
	
	As shown in Fig. \ref{fg:Gt_Jw}(a), the mono-exponential decay in $G_{SBM}(t)$ is not consistent with the multi-exponential (i.e., stretched) decay in $G(t)$. Equivalently, $J(\omega)$ in Fig. \ref{fg:Gt_Jw}(b) does not follow the $f^{-2}$ power-law behavior at large $f$. This is expected since the simulations implicitly include both inner-sphere and outer-sphere (see below) contributions. Currently the simulations do not separate inner-sphere from outer-sphere contributions, therefore the simulations do not clarify the origin of the multi-exponential decay in $G(t)$.
	
	It is also possible that the inner-sphere $G(t)$ itself has a multi-exponential decay of its own. The functional form of $G_{SBM}(t)$ is based on the Debye model, which was previously shown to be inaccurate when used in the BPP model for liquid alkanes and water \cite{singer:jmr2017}. It would therefore not be surprising if the inner-sphere $G(t)$ was also multi-exponential in nature.
	
	Assuming that inner-sphere relaxation dominates, and therefore that the multi-exponential decay in $G(t)$ is due to inner-sphere dynamics alone, the correlation time $\left<\tau\right> \simeq$ 30 ps is consistent with published values from the SBM inner-sphere model, where a range of $\tau_R \simeq 23 \leftrightarrow 45$ ps is reported \cite{koenig:jcp1975,banci:1985,powell:jacs1996,strandberg:jmr1996,rast:jcp01,lindgren:pccp2009,luchinat:jbio2014}. Note that $\left<\tau\right> \simeq$ 30 ps is a factor $\simeq 10$ larger than that of bulk water ($\tau_R \simeq 2.7$ ps  \cite{singer:jmr2017}), which is expected given the hindered rotational motion of the Gd$^{3+}$--aqua complex. 
	
	Continuing with the assumption that inner-sphere relaxation dominates, one can approximate the following inner-sphere quantities:
	\begin{align}\frac{1}{r_{in}^6} &\simeq \frac{1}{2q}\sum_{i=1}^N
		\left< \! \frac{1}{r_{i}^6(t')} \!\right>_{\!\! t'} \label{eq:SBM_R} \\
		\Delta\omega^2_{in} &\simeq \frac{1}{2q}\sum_{i=1}^N\Delta\omega_i^2
		\label{eq:SBM_SM} \\
		\frac{1}{T_{1in}} &\simeq \frac{1}{2q}\sum_{i=1}^N\frac{1}{T_{1i}}.
		\label{eq:SBM_T1} \\
	\end{align}
	The expressions are averaged over the $2q$ $^1$H nuclei in the inner sphere, where $q = 8.5$ is the H$_2$O inner-sphere coordination number determined from $n(r)$ (Fig. \ref{fg:Gr_Nr}) at $r \simeq 3.5$ {\AA}. Note again that these equations neglect the outer-sphere contribution, implying that $\Delta\omega_{in}^2$ is an upper bound, while $r_{in}$ and $T_{1in}(0) $ are lower bounds.
	
	According to Table \ref{tb:MD}, the resulting $r_{in} \simeq $ 2.97 {\AA}  is consistent with published values from the inner-sphere SBM model, where a range of $r_{in} \simeq 3.0 \leftrightarrow 3.2$ {\AA} is reported \cite{koenig:jcp1975,banci:1985,powell:jacs1996,strandberg:jmr1996,rast:jcp01,lindgren:pccp2009,luchinat:jbio2014}.
	The product $\Delta\omega_{in} \left<\tau\right> \simeq 0.002 $ indicates that the Redfield-Bloch-Wagness condition ($\Delta\omega \left<\tau\right>  \ll 1$) is satisfied \cite{abragam:book,mcconnell:book,cowan:book}, which justifies the relaxivity analysis used here. The fast-exchange regime \cite{korringa:pr1962} can also be verified by noting that $T_{1in}(0) \simeq$ 4.4 $\mu$s at $f_0 = 0$ is three orders of magnitude larger than $\tau_m \simeq 1$ ns, i.e. $(T_{1in} + \tau_m) \simeq T_{1in} $ can be assumed.

	\subsection{HF outer-sphere model}\label{subsc:HF}
	
	We now discuss the outer-sphere contribution to relaxivity, although it is generally believed (though not proven) to be smaller than inner-sphere relaxivity \cite{koenig:jcp1975,lauffer:cr1987,caravan:cr1999}. The outer-sphere contribution is expected to follow the Hwang-Freed (HF) model for the relative translational diffusion between Gd$^{3+}$ and H$_2$O assuming two force-free hard-spheres \cite{hwang:JCP1975}:
	\begin{multline}
		G_{HF}(t)=G_{HF}(0)\frac{54}{\pi}\int\limits_0^\infty  \frac{x^2}{81 + 9 x^2 - 2 x^4 + x^6} \times \\ \exp\left(-x^2 \frac{t}{\tau_D} \right)  dx. \label{eq:HS2}
	\end{multline}
	The translational-diffusion correlation time $\tau_D$ is defined as the average time it takes the molecule to diffuse by one hard-core diameter $d$. $G_{HF}(t)$ is a multi-exponential decay by nature, and therefore one expects the total $G(t)$ to be stretched, the extent of which depends on the relative contributions of outer-sphere to inner-sphere.
	
	The correlation time $\tau_D$ can be predicted as such:
	\begin{align}\tau_D &=  \frac{d^2}{D_{W} + D_{Gd}} = \frac{9}{4}\tau_T, \label{eq:HS3}
	\end{align}
	where the simulated diffusion coefficients of H$_2$O ($D_W$) and Gd$^{3+}$ ($D_{Gd}$) in Gd$^{3+}$--aqua are taken from Fig. \ref{fg:Diff}, the results of which are listed in Table \ref{tb:Deff}. The hard-core diameter $d$ is taken from the local maximum at $r \simeq 4.6 $ {\AA} in the pair-correlation function $g(r)$ in Fig. \ref{fg:Gr_Nr} (which is attributed to outer-sphere water). Note that the resulting $\tau_D \simeq 99$ ps is a factor $\simeq 10$ larger than that of bulk water ($\tau_D \simeq 9.0$ ps \cite{singer:jmr2017}), which is expected given the larger hard-core distance $d \simeq 4.6$ {\AA} than bulk water ($d \simeq $ 2.0 {\AA} \cite{singer:jmr2017}). 
	
	\begin{table}[!ht]
		\centering
		\begin{tabular}{ccc|cc}				
			\hline
			$^{{\strut}{}}$	$D_{W}$& $D_{Gd}$ & $d$ &$\tau_D$ & $\tau_T$ \\
			$^{{\strut}{}}$	(${\rm \AA}^2$/ps) & (${\rm \AA}^2$/ps) & (${\rm \AA}$) & (ps) & (ps)   \\
			\hline
			$^{{\strut}{}}$	 0.19   & 0.03 & 4.6 & 99 & 44  
		\end{tabular}
		\caption{Diffusion coefficients of H$_2$O ($D_W$) and Gd$^{3+}$ ($D_{Gd}$) in Gd$^{3+}$--aqua, distance of closest approach ($d$) between Gd$^{3+}$ and outer-sphere H$_2$O according to $g(r)$ (Fig. \ref{fg:Gr_Nr}), and, resulting  translational-diffusion correlation time $\tau_D \,(= 9/4\,\tau_T)$ from Eq. \ref{eq:HS3}.} \label{tb:Deff}
	\end{table}
	
	The value $\tau_D \simeq 99$ ps is compared to the distribution $P(\tau)$ in Fig. \ref{fg:Ptau}. More specifically, the translational correlation time $\tau_T \simeq 44$ ps is plotted in Fig. \ref{fg:Ptau}, where the relation $\tau_D = 9/4 \,\tau_T$ and the origin of the factor 9/4 is explained in \cite{cowan:book,singer:jcp2018}. $\tau_T $ lies within the $P(\tau)$ distribution, indicating that the outer-sphere may contribute to relaxivity. Further investigations are underway to separate inner-sphere from outer-sphere contributions in the simulations, without assuming any models.

	We note that $P(\tau)$ in Fig. \ref{fg:Ptau} has a small contribution at short $\tau\simeq 10^{-1}$ ps, which is a result of the sharp drop in $G(t)$ over the initial $t \simeq 0.2 $ ps (Fig. \ref{fg:Gt_Jw}(a)). This molecular mode is also present for intramolecular relaxation in liquid alkanes and water, while it is absent for intermolecular relaxation. In the case of alkanes, the ubiquitous intramolecular mode at $\tau\simeq 10^{-1}$ ps is attributed to the fast spinning methyl end-groups \cite{singer:jcp2018}. Investigations are underway to better understand the origin of this mode in Gd$^{3+}$--aqua, which can perhaps be explained using a two rotational-diffusion model such as found in bulk water \cite{madhavi:jpcb2017}. The origin of other modes in $P(\tau)$ at $\tau\simeq 10^{0}$ ps and $\simeq 10^{1}$ ps are also being investigated.
	
	Finally, we note that the $r_{1}$ dispersion in Fig. \ref{fg:R1_relaxivity} results in a mild increase in the ratio $T_{1}/T_{2} = r_{2}/r_{1} \simeq 7/6$ above $f_0 \gtrsim 10$ MHz, until $f_0 \gtrsim 6$ GHz where $T_{1}/T_{2} $ increases further. Combining Eqs. \ref{eq:R11} and \ref{eq:R12} with $J(\omega_e) = 0$ (i.e., slow-motion regime) and $J(\omega_0) = J(0)$ (i.e., fast-motion regime) accounts for the factor $T_{1}/T_{2} = $ 7/6 within the frequency range $f_0 \simeq$ 10 MHz $\leftrightarrow$ 6 GHz. The ratio $T_{1}/T_{2} = 7/6$ was also used to explain water-saturated sandstones \cite{foley:JMR1996}.

	\subsection{Electron-spin relaxation} \label{subsc:ESR}
	
	At low-frequencies ($f_0 \lesssim 0.5$ MHz), the difference between $r_1$ measurements ($\simeq$ 29.7 mM$^{-1}$s$^{-1}$) and simulation ($\simeq$ 34.6 mM$^{-1}$s$^{-1}$) can be reconciled by taking the zero-field electron-spin relaxation time $T_{e0} = T_{1e}(0) = T_{2e}(0)$ into account. Assuming that the correlation times $P(\tau)$ are uncorrelated with the electron-spin relaxation times, the following expression results \cite{bloembergen:jcp1961}: 
	\begin{align}
		r_{1,2}'(0) & =\frac{1}{[H]}\frac{20}{9} \Delta\omega^2\!\left<\tau'\right>, \label{eq:EL1}\\
		\frac{1}{\left<\tau'\right>}  &=\frac{1}{\left<\tau\right>} +\frac{1}{T_{e0}}.\label{eq:EL2}
	\end{align}
	This is equivalent to introducing an exponential decay term $\exp(-t/T_{e0})$ inside the FFT integral of Eq. \ref{eq:R10}. The fitted value of $T_{e0} \simeq 180 $ ps is determined by matching $r_1'(0)$ to the average low-frequency ($f_0 \lesssim 0.5$ MHz) measurement (Fig. \ref{fg:R1_relaxivity}). The resulting $T_{e0}\simeq 180 $ ps is consistent with the published range of $T_{e0} \simeq 96 \leftrightarrow 160 $ ps 
	\cite{koenig:jcp1975,banci:1985,powell:jacs1996,strandberg:jmr1996,rast:jcp01,lindgren:pccp2009,luchinat:jbio2014}. Investigations are underway to incorporate  $T_{1e}(\omega_e)$ and $T_{2e}(\omega_e)$ dispersion  \cite{kowalewski:jcp1985,rast:jcp01,borel:jacs2002,lindgren:pccp2009} for predicting $r_1$ at low frequencies.
	
	\section{Conclusions}\label{sc:Conclusions}
	
	Atomistic MD simulations of $^1$H NMR relaxivity $r_1$ for water in Gd$^{3+}$--aqua complex at 25$^{\circ}$C show good agreement (within $\simeq $ 8\%) with measurements in the range $f_0 \simeq $ 5 $\leftrightarrow$ 500 MHz, without any adjustable parameters in the interpretation of the simulations, and without any relaxation models. This level of agreement validates the simulation techniques and analysis of Gd$^{3+}$--$^1$H dipole-dipole relaxation for unlike spins. The simulations show potential for predicting $r_1$ at high frequencies in chelated Gd$^{3+}$ contrast-agents for clinical MRI, or at the very least the simulations could reduce the number of free parameters in the Solomon-Bloembergen-Morgan (SBM) inner-sphere and Hwang-Freed (HF) outer-sphere relaxation models.
	
	Simulations suggest residence times between $\tau_m \simeq 1 \leftrightarrow$ 2 ns for a complete rejuvenation of the inner sphere waters of Gd$^{3+}$. Further, the average coordination number is $q \simeq 8.5$. These observations are consistent with previously reported interpretation of experiments using the SBM model.
	
	The autocorrelation function $G(t)$ shows a multi-exponential decay, with an average correlation time of $\left<\tau\right>\simeq$ 30 ps. The multi-exponential nature of $G(t)$ is expected given that the simulation implicitly includes both inner-sphere and outer-sphere contributions. The results are analyzed assuming that the inner-sphere relaxation dominates, yielding approximations for the average Gd$^{3+}$--$^1$H separation $r_{in} \simeq 2.97$ {\AA} and rotational correlation time $\tau_R = \left<\tau\right>\simeq$ 30 ps, both of which are consistent with previously published values which use the SBM model. 
	
	A distance of closest approach (i.e. hard-core diameter) for Gd$^{3+}$--H$_2$O of $d\simeq 4.64$ {\AA} is determined from the local maximum in $g(r)$ (attribute to outer-sphere water), which together with the simulated diffusion coefficients of Gd$^{3+}$ and H$_2$O are used to estimate the translational-diffusion correlation time $\tau_D = 9/4 \,\tau_T \simeq 99$ ps in the HF outer-sphere model. Comparing $\tau_T$ to the distribution in molecular modes $P(\tau)$ (determined from the modes expansion of $G(t)$) indicates that the outer-sphere may contribute to relaxivity. Further investigations are underway to separate inner-sphere from outer-sphere contributions in the simulations, without assuming any models. 
	
	Below $f_0 \lesssim $ 5 MHz the simulation overestimates $r_1$ compared to measurements, which is used to estimate the zero-field electron-spin relaxation time $T_{e0} $. The resulting fitted value $T_{e0} \simeq 180 $ ps is consistent with the published range of values. Further investigations are underway to incorporate dispersion in the electron-spin relaxation time.
	
	\section*{Acknowledgments} \label{sc:Acknow}
	We thank Vinegar Technologies LLC, Chevron Energy Technology Company, and the Rice University Consortium on Processes in Porous Media for financial support. We gratefully acknowledge the National Energy Research Scientific Computing Center, which is supported by the Office of Science of the U.S. Department of Energy (No.\ DE-AC02-05CH11231) and the Texas Advanced Computing Center (TACC) at The University of Texas at Austin for high-performance computer time and support. 
	

%

\end{document}